\documentclass[superscriptaddress,reprint,twocolumn,amsmath,amssymb, prl]{revtex4}
\pdfoutput=1
\usepackage{CJK}
\usepackage{natbib}
\usepackage{amsmath}
\usepackage{amssymb}
\usepackage{epsfig}
\usepackage{graphics}
\usepackage{xcolor}

\graphicspath{{figures/}}

\newcommand{\un}[1]{\ensuremath{\,{\rm #1}}}
\newcommand{\sci}[2]{\ensuremath{#1{\times}10^{#2}}}

\begin{document}

\title{First measurements of high frequency cross-spectra from a pair of large Michelson interferometers}

\author{Aaron S. Chou}
\affiliation{Fermi National Accelerator Laboratory} 
\author{Richard Gustafson}
\affiliation{University of Michigan} 
\author{Craig Hogan}
\affiliation{Fermi National Accelerator Laboratory}
\affiliation{University of Chicago}
\author{Brittany  Kamai}
\affiliation{University of Chicago}
\affiliation{Vanderbilt University}
\author{Ohkyung Kwon}
\affiliation{University of Chicago}
\affiliation{Korea Advanced Institute of Science and Technology (KAIST)}
\author{Robert Lanza}
\affiliation{University of Chicago}
\affiliation{Massachusetts Institute of Technology}
\author{Lee McCuller}
\affiliation{University of Chicago}
\affiliation{Massachusetts Institute of Technology}
\author{Stephan S. Meyer}
\affiliation{University of Chicago}
\author{Jonathan Richardson}
\affiliation{University of Chicago}
\author{Chris Stoughton}
\affiliation{Fermi National Accelerator Laboratory}
\author{Raymond Tomlin}
\affiliation{Fermi National Accelerator Laboratory}
\author{Samuel Waldman} 
\affiliation{SpaceX}
\author{Rainer Weiss}
\affiliation{Massachusetts Institute of Technology}

\begin{abstract}
Measurements are reported of the cross-correlation of spectra of differential
position signals from the Fermilab Holometer, a pair of co-located 39~m long,
high power Michelson interferometers with flat, broadband frequency response in
the MHz range.   The instrument obtains sensitivity to high frequency correlated
signals far exceeding any previous measurement in a broad frequency band
extending beyond the 3.8~MHz inverse light crossing time of the apparatus.  The
dominant but uncorrelated shot noise is averaged down over $2\times 10^8$
independent spectral measurements with 381~Hz frequency resolution to obtain
$2.1\times 10^{-20} \ \mathrm{m}/\sqrt{\mathrm{Hz}}$ sensitivity to stationary
signals.  For signal bandwidths $\Delta f > 11$~kHz, the sensitivity to strain
$h$ or shear power spectral density of classical or exotic origin surpasses a
milestone $PSD_{\delta h} < t_p$ where $t_p= 5.39\times 10^{-44}/\mathrm{Hz}$ is
the Planck time.
\end{abstract}
\pacs{??}

\maketitle

In this Letter, we report first data from a pair of co-located and co-aligned
39.06~m long power-recycled Michelson interferometers, each operating at 2~kW
power with mean shot-noise-limited differential position noise sensitivity of
$2.1\times 10^{-18} \ \mathrm{m}/\sqrt{\mathrm{Hz}}$. The apparatus adopts many
of the technologies developed for sub-kHz gravitational wave detection
\cite{Weiss1972,Saulson1994,LIGO2009,Adhikari2014,2007OptCo.280..492W,Affeldt:2014rza},
but is instead optimized for a much larger signal bandwidth extending up to
25~MHz. Whereas gravitational wave interferometers incorporate Fabry-Perot arm
cavities and/or an output port recycling cavity to resonantly enhance the
instrumental response to differential strain at low frequencies, the new
interferometers employ only common mode power recycling cavities and thus do not
low-pass filter the differential signals. They thus maintain their full
Michelson differential bandwidth at frequencies up to the 3.8~MHz inverse
light-crossing time of the apparatus~\cite{Rakhmanov:2008is}. The signal
fluctuations in interference fringe power are digitized at 50~MHz to achieve a
detection bandwidth much larger than that of typical gravitational wave
detectors. While some prototype resonant detectors have been operated at such
high frequencies~\cite{Bernard:2001kp,Cruise:2006zt,Akutsu:2008zza}, the
broadband strain sensitivity of the new instrument far exceeds that of
previously reported narrowband results at these frequencies. Moreover, similarly
to GEO600~\cite{Affeldt:2014rza,Dooley:2014nga} but unlike
LIGO~\cite{Adhikari2014}, the full recycled laser power of each new
interferometer is incident on the beam splitter, thus giving equal sensitivity
to longitudinal strain and to transverse shear fluctuations, as measured
relative to the laser beam propagation direction in each interferometer arm.

A large improvement in sensitivity to external but local stationary signals
common to both interferometers is achieved by cross-correlating the outputs of
the two devices to average away uncorrelated noise. While this interferometer
cross-correlation technique has been demonstrated by the LIGO Scientific
Collaboration~\cite{Abbott:2009ws,Abadie:2011fx,Aasi:2014zwg,Aasi:2014jkh}, the
higher frequency signal band of the Holometer enables two significant
improvements in noise reduction. First, while LIGO has operated two co-located
interferometers H1 and H2 in the same vacuum system at the Hanford site, the
cross-correlation analysis has been complicated by substantial contributions of
correlated environmental noise at low frequencies. In contrast, the $f\gtrsim
1$~MHz signal band of the new instrument is largely free of this low frequency
seismic and acoustic noise. Secondly, the enhanced signal bandwidth, as large as
$\Delta f = 25$~MHz (compared to the $<1$~kHz bandwidth of typical gravitational
wave detectors) reduces the time required per independent measurement and thus
enables a much larger noise averaging factor $\sqrt{N_\mathrm{meas}} =
\sqrt{\tau_\mathrm{int} \cdot \Delta f}$ for any cumulative integration time
$\tau_\mathrm{int}$. In an example described below, the new spectral data are
analyzed to test a speculative model of Planckian diffraction, and the noise is
averaged down by a factor of $\sqrt{(145 \ \mathrm{hours})\cdot(700
\ \mathrm{kHz})} \approx 6\times 10^{5}$. The data constrain strain or shear power
spectral density in the detection band to be $PSD_{\delta h} < 0.25\times t_p$
where $t_p= 5.39\times 10^{-44}/\mathrm{Hz}$ is the Planck time.

\noindent {\bf Experimental design} --- In each interferometer, continuous wave
$\lambda= 1064\un{nm}$ laser light is injected to a beamsplitter, divided into
two orthogonal arms and reflected at distant end mirrors. The returning beams
coherently interfere at the beamsplitter, with intensity varying as $P_{\rm
fringe} = P_{\rm BS}(\epsilon_\text{cd} + (1-2\epsilon_\text{cd})\sin^2(2\pi
X/\lambda) )$ at the antisymmetric port. In this expression, the differential
arm length (DARM) is given by $X \equiv L_1-L_2$ where $L_1 \approx L_2 =
39.06$~m are the lengths of the two arms. Perturbations $\delta X$ due to either
strain or shear are imprinted as amplitude modulation on the output fringe
power. $P_{\rm BS}$ is the power incident on the beamsplitter and the contrast
defect parameter $\epsilon_\text{cd}$ characterizes residual leakage of
non-interfering light caused by geometrical mismatches in the beams returning
from the two arms.

The remaining power exiting the symmetric beam splitter port and returning
towards the laser is instead reflected back into the device using a 1000~ppm
transmission mirror. The insertion of this input coupling mirror forms an
overcoupled Fabry-Perot cavity with free spectral range $\mathrm{FSR} \approx
3.8$~MHz determined by the common arm length $(L_1+L_2)/2$. The laser is
frequency-locked to the instantaneous cavity frequency via the Pound-Drever-Hall
(PDH) technique \citep{DreverHall1983,black00} to achieve a typical power
build-up from the injected 1.1~W laser power to intracavity power $P_{\rm BS}
\approx 2$~kW into the beamsplitter from the recycling mirror. The 900~Hz
transmission bandwidth of the optical cavity filters higher frequency amplitude
and phase noise present on the incident laser beam. It also indicates a total of
$\approx 1470$ PPM round-trip loss including recycling transmissivity,
scattering/absorption losses, Michelson fringe offset and defect leakage.

\begin{figure}[ht]
  \centering
  \includegraphics[width=\linewidth]{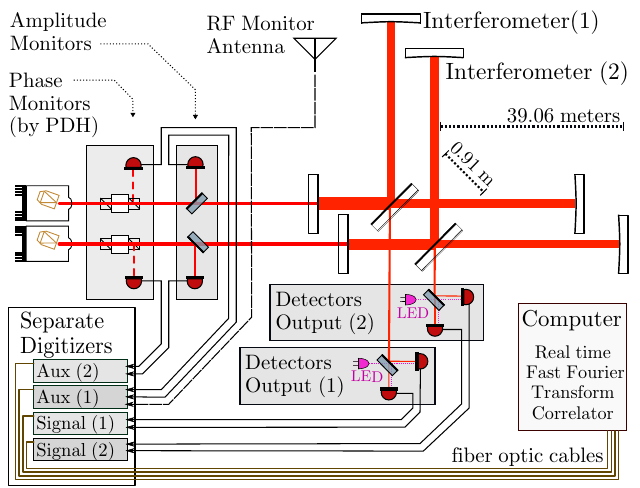}
  \caption{ Schematic of the two co-located interferometers and associated data
acquisition channels. The two devices are optically and electrically isolated to
eliminate cross-talk.}
  \label{fig:instrument}
\end{figure}

To produce a linear response to differential length perturbations $\delta X$,
each interferometer is operated at a DARM offset of around 1~nm from a dark
fringe. No optic is suspended, but steel and Viton spring stacks damp
environmental noise above 15Hz. A digital control system monitors fluctuations
in the output light and feeds back differential signals to piezo-electric
actuated end mirror mounts to hold the fringe offset within the 700Hz unity gain
frequency. The shaped loop maintains a $50$ pm RMS residual motion as measured
the through the 16kHz control system Nyquist frequency. Narrowband excitations
used for length sensitivity calibration and alignment control account for half
of this RMS. At this fringe offset, around 50~ppm of signal-bearing interference
light appears at the antisymmetric port, as measured relative to the intracavity
power. This value is chosen to balance the interference fringe light with the
non-interfering contrast defect light leakage $\epsilon_\text{cd} \approx
50\un{ppm}$ which carries no signal but contributes shot noise variance.

Accounting for detection inefficiencies and contrast defect degradation, the
mean shot-noise-limited displacement sensitivity due to the 2~kW power incident on
the beamsplitters is $\mathrm{PSD}^{\text{shot}}_{\delta X} \approx
(\sci{2.1}{-18}\un{m/\sqrt{Hz}})^2$; The excess power spectral density compared
to an ideal 2 kW Michelson is $2.1\times$ in one machine and $3.5\times$ in the
other. These sensitivities are confirmed by calibration measurements as
summarized below. The signals from the two interferometers are cross-correlated
and averaged to reach sensitivity more than four orders of magnitude below the
shot noise limit.

The success of the cross-correlation averaging technique depends on low
instrumental correlation between the two interferometers and requires nearly
complete independence of the two devices, despite sharing an experimental hall.
Each interferometer is enclosed in its own vacuum system, and the injection and
control systems are operated on separate optics tables and electronics racks.
The digitizers for the two instruments are isolated and independently
synchronized to GPS, communicating with the realtime spectrum processing
computer only through optical fiber (see Fig.~\ref{fig:instrument}).

\noindent {\bf Data acquisition and methodology} --- 200 mW of detected
asymmetric port power in each interferometer is required to balance between
photodiode limits and excess noise from contrast defect. The large dynamic range
between this DC power and the shot noise level presents challenges for linear
detection. The output power is split by a secondary beam splitter to divide it
between two custom low transimpedance photoreceivers based on high linearity,
2~mm InGaAs photodiodes. This linearity is demonstrated in each detector with DC
power up to 150 mW. A low gain DC amplification channel samples the photocurrent
and has flat response from DC-80~kHz. A high gain, transimpedance-based,
AC-coupled radiofrequency (RF) channel is best calibrated between 900~kHz-6~MHz
and digitizes to a 25~MHz Nyquist frequency. Outside this band, the phase
matching of the calibration deteriorates, impacting the real projection of the
complex cross correlation.

The two interferometers together thus have four RF output streams, each
digitized at 100 MHz sample rate with 14 bits, and then downsampled to 50 MHz.
These four channels along with an additional four auxiliary monitor channels are
Fast Fourier Transformed in real time with 381~Hz frequency resolution. A
symmetric $8\times 8$ cross-spectrum matrix is computed. The realtime
computation rate limits the digitization to 50 MHz. Measurements of
these 36 (cross-)spectra are averaged over 700 sequential spectral measurements
(around 1.8~s) for storage. The remaining averaging is performed in offline
analysis.

Isolation of the two interferometers is established by measurement, as
described below in the backgrounds section. During the accumulation for this
result the auxiliary channels are set at various times to monitor the PDH laser
phase noise, the laser intensity noise, and loop antennas detecting the local RF
environment.

\noindent {\bf Absolute calibration of sensitivity} --- An indirect calibration
ladder is used to establish the instantaneous length sensitivity at MHz
frequencies. Because of resonances in the piezo stacks actuating the end
mirrors, a mechanical dither signal can only be injected at a low frequency of
1~kHz, whereas the RF detector channel is high-passed at 900~kHz. The 1~kHz
dither is calibrated by misaligning the cavity mirrors to operate the
interferometers in a non-power-recycled configuration with a simple Michelson
response. The end mirrors are then slowly actuated to sweep across an entire
interference fringe to reference the voltage signal to the 1064~nm wavelength.
After correcting for the measured interferometer control system feedback, the in situ
dither amplitude is determined to be $10^{-11}$~m. Measurements of the
low-passed DC and the high-passed RF transfer functions of the photoreceivers
refer the 1~kHz length-calibrated in situ dither to the signal band above 1~MHz
with 5\% systematic uncertainty. The resulting calibration matches the
sensitivity expected from fitted interferometer parameters of cavity power,
contrast defect and the DARM fringe offset, fit using slow controls data of the
calibration lines and readouts of the asymmetric port, arm transmission
power and cavity reflection power. 

\noindent {\bf In situ monitoring of data quality} -- During data-taking
operations, the 1 kHz DARM dither is run continuously. For each detector, both
the DC photocurrent and the 1 kHz signal are monitored from the detector DC
channel and the ratio of these measures is a proxy for the instantaneous fringe
offset. The shot noise level in the 1-2~MHz signal band is also continuously
monitored and the ratio of this to the DC photocurrent signal monitors the
relative stability of the photoreceiver RF and DC channel responses. These and
other observables such as the power reflected from the cavity back towards the
laser and the power transmitted through the end mirrors serve to monitor the
stability of the calibrated sensitivity of the instrument to position
disturbances. The uncertainty in calibration from both systematic uncertainties
and run-to-run variability is less than 10\%.

Periods with abnormal operating conditions are vetoed prior to accumulation into
the averaged spectra. To verify the control system lock to a stable fringe
offset, the low frequency photocurrent is continuously monitored and periods of
lock loss are rejected. Periods of enhanced RF noise exceeding shot noise by
20\% are also rejected. Fast noise glitches are identified by a threshold veto
on the raw time-series photocurrent data, preventing ADC clipping. During
transition periods when the control system lock of the interferometer is lost or
is being reacquired, 4 seconds of data immediately before the lock loss and
immediately after a lock reacquisition are vetoed. During active data-taking,
the duty cycle for stable operations is greater than 80\%.

To monitor the timing stability of the cross-correlation data acquisition
system, separate LED flashers driven from a common source inject a 13MHz timing
calibration line to the output photodiodes. The LED signal amplitude
and phase coherence in each detector is continuously recorded and indicates that
the electronically isolated digitizers have high phase stability for frequencies
up to 25~MHz.

\begin{figure}[t]
  \centering
  \includegraphics[width=\linewidth]{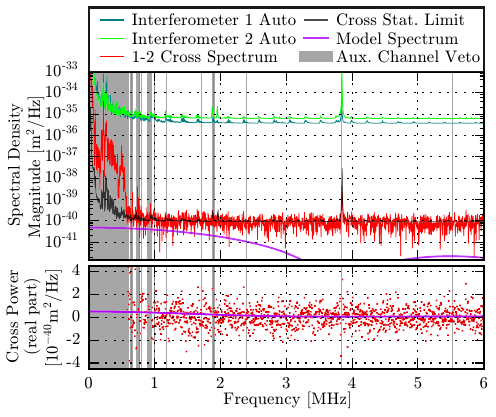}
  \caption{ 
  Accumulated power spectra with 3.81 kHz resolution, zoomed in to frequencies
  near the free spectral range.   In the upper panel, the two upper curves show
  the output PSD (averaged over the two photodetectors) for each interferometer.
  Below that are two curves showing the expected noise in the cross correlation
  based on $\sqrt{N_\mathrm{spectra}}$ averaging of the PSDs, and the observed
  magnitude of the cross correlation data.  The lower panel plot is on a linear
  scale and is the real part of the cross correlation spectrum.  This is the
  signal integrated in Fig.~\ref{fig:cumulative_test}. Both panels also show an
  example broadband model spectrum with Planckian normalization as given by
  Eq.~\ref{E:holo_model}.
  }
  \label{fig:power_spectrum}
\end{figure}

\noindent {\bf Measured spectra} --- Fig.~\ref{fig:power_spectrum} shows the
measured auto- and cross-spectra averaged over 145~hours of data taken in
July-August, 2015. The auto-spectrum for each individual interferometer is
obtained by a weighted average of its two output photodetectors, with weighting
given by the instantaneous calibrated DARM sensitivity. The many subsequent
measurements are also similarly weighted when summed into the average. The raw
381 Hz resolution spectra are frequency-averaged to

produce spectra with 3.81~kHz resolution and negligible bin-to-bin correlation.
At high frequencies, these spectra are shot-noise-limited as expected with flat
regions well described by Gaussian noise. A repeating sequence of peaks is due
to thermally excited acoustic modes of individual optics substrates. The
magnitude of the resolved acoustic lines is consistent with that expected from
the ambient temperature. Excess power is also seen at higher order mode
resonances of the Fabry-Perot cavity for each interferometer and at the 3.8~MHz
FSR. At these resonances, amplitude and phase noise of the laser is no longer
efficiently filtered by the cavity, leaking through to the output port. Because
the interferometers use independent optics and lasers, the excess noise from
these sources is uncorrelated but reduces the sensitivity of the experiment at
affected frequencies.

The measured cross-spectral data are projected onto the real axis to search for
correlation at zero time delay. The shot-noise-limited measured power is
consistent with the expected statistical sensitivity with
$\sqrt{N_\mathrm{spectra}}$ improvement from averaging. The data are verified to
be normally distributed with no statistically significant outliers.

\noindent {\bf Backgrounds and frequency bin vetoes} --- A limited set of
potential backgrounds is studied in order to constrain the possible destructive
interference of environmental contamination with a putative signal spectrum. The
laser phase and amplitude noise spectra are measured in situ via optical
pick-offs prior to injection, and recorded in the auxiliary RF channels. The
cross-spectra of these channels with the interferometer output channels is
calibrated using ex-situ transfer function measurements. At frequencies below 1
MHz, the interferometer output spectra are dominated the $1/f$ laser phase
noise, incompletely suppressed by the cavity filter. Frequency bins with high
coherence to the laser phase and amplitude monitors of the opposite
interferometer, or with external antenna channels, are vetoed for the analysis.
Data below 100~kHz are vetoed due to a large environmental noise component,
while the auxiliary channels enforce vetoes at frequencies up to 600kHz and
sporadically above that. These vetoes rely only on auxiliary channels and do not
systematically bias the search for signal power in the interferometer output.
Vetoed regions are shaded in gray in the plots. For remaining bins, correlated
or anti-correlated laser noise is statistically limited to be $<3\%$ of the
estimated Planckian power spectrum. Furthermore, dark noise studies indicate
correlated electronic pickup and environmental light (including the LED timing
calibrators) to be $<1\%$ of the statistical sensitivity.

\begin{figure}[t]
  \centering
  \includegraphics[width=\linewidth]{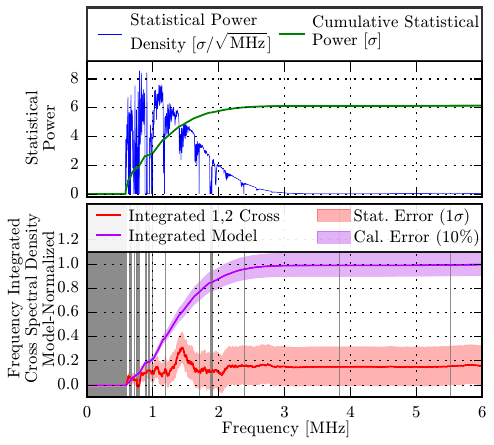}
  \caption{ 
The upper panel shows the predicted measurement significance (model
signal/instrument noise) in each frequency bin as well as its integral.
The lower panel shows the frequency integrated measurement with a shaded
$1\sigma$ uncertainty limit along with the integrated model spectrum.  Both are
normalized to the predicted model amplitude. The shaded region around the model
curve is the 10\% calibration uncertainty. The grey bands are frequencies
vetoed using ancillary and housekeeping data.   }
  \label{fig:cumulative_test}
\end{figure}

\noindent {\bf Model testing} --- As an example of how the spectral data can be
used, we consider a speculative model in which irreducible space-time noise
arising from a putative fundamental Nyquist frequency $f_p = 1/t_p$ grows via
diffraction over macroscopic distances to give a white noise shear power
spectral density quantitatively equal to $t_p$.
In units of position variance, the predicted power spectral density (plotted as
the purple curve in Fig.~\ref{fig:power_spectrum}) takes the following
representative form~\cite{kwon:2014yea}:
\begin{equation}
    \mathrm{PSD}_{\delta X}(f) = \frac{t_p} {\sqrt{\pi}} \cdot L^2 \cdot  \frac{\sin^2({\pi  (2L/c)f})}{({\pi  (2L/c)f})^2} \label{E:holo_model} \\  
\end{equation} 
which is a sinc response function normalized to $\sci{4.64}{-41} \un{m}^2 /
\un{Hz}$ and distributed over the $3.8$~MHz free spectral range for arm length
$L\approx 39$~m. While this space-time noise is expected to be correlated
between co-located interferometers, the decoherence due the small separation
$d=0.91$~m between the two beam splitters would cause a $d/L=2.3\%$ reduction in
the normalization of the model cross-spectrum.

To optimize sensitivity to the predicted spectral shape, each non-vetoed
3.81~kHz bin is weighted by the (predicted) signal-to-noise ratio with signal
estimated from the model spectrum Eq.~\ref{E:holo_model} and noise variance
estimated from the measured interferometer auto-spectra divided by $\sqrt{N_{\rm
spectra}}$. Fig.~\ref{fig:cumulative_test} shows this result in the form of a
weighted frequency integral of the cross-spectrum data from
Fig.~\ref{fig:power_spectrum}. Plotted on the upper panel of
Fig.~\ref{fig:cumulative_test} is the measurement weight shown as a potential
signal significance density for each frequency bin ($\sigma/\sqrt{\un{MHz}}$).
As discussed above, the shot noise is exceeded at some frequencies by other
uncorrelated stochastic noise sources, causing dips in the expected significance
density which reduce the instrument's integrated sensitivity by about 10\% while
causing no systematic bias. For this model, about $20\%$ of the potential signal
significance comes from frequencies below 1 MHz, $70\%$ from frequencies between
1 and 2 MHz, and $10\%$ from above 2 MHz. The integrated significance shows the
potential for $6.2\sigma$ statistical sensitivity for detecting or rejecting
this model.

The lower panel of Fig.~\ref{fig:cumulative_test} shows the frequency integral
of the predicted signal given in equation~\ref{E:holo_model} weighted by the
expected signal to noise ratio. This curve is normalized to integrate to unity
with a shaded band representing the 10\% calibration uncertainty. The lower
curve in this plot is the corresponding integral of the weighted data points.
This curve exhibits a random walk, thus indicating that the significance
accumulation has no excessive contribution from any particular frequency band.
The integral takes into account the small correlations between adjacent
frequency bins due to apodization and sampling. The shaded vertical bands are
vetoed regions as described above. The endpoint to the right of the plot is the
total integrated signal and is the result of this analysis. The shaded band
around the curve is the $\pm 1\sigma$ accumulated statistical uncertainty.

Using all data up to 25~MHz, the weighted integral curve remains statistically
consistent at $1.1\sigma$ with zero broadband correlation. The model of
Eq.~\ref{E:holo_model} is thus excluded with $5.1\sigma$ statistical
significance, reduced by the 10\% calibration uncertainty to $4.6\sigma$.
Alternatively, the result may be viewed as a constraint on the normalization of
this model to be less than 44\% of the predicted value at 95\% confidence level.
It should be emphasized that these results apply only to the spectral shape of
the particular model used here. Similar analysis techniques should be used for
testing of any model which predicts shear or strain variance power in this
detection band.

\noindent {\bf Conclusions} --- Modern interferometers including the ones
described here are now achieving correlated strain sensitivity surpassing
Planckian normalization, and thus may provide data useful for searching for new
effects potentially arising from Planck scale microphysics. Further studies will
survey with improved sensitivity other potential models with possible Planckian
information content accessible to the current
instrument\cite{Hogan:2015kva}. These measurements will also provide uniquely
deep constraints on gravitational waves in the MHz band. While the apparatus in
its current Michelson layout is equally sensitive to shear and strain noise, it
would not respond to correlated exotic noise power in rotational observables;
these could be studied with a similar instrument reconfigured with bent
arms~\cite{Hogan:2015sra}.
 
\begin{acknowledgments}
This work was supported by the Department of Energy at
Fermilab under Contract No. DE-AC02-07CH11359 and the Early Career Research
program (FNAL FWP 11-03), and by grants from the John Templeton Foundation; the
National Science Foundation (PHY-1205254, DGE-0909667, DGE-0638477,
DGE-1144082), NASA (NNX09AR38G), the Fermi Research Alliance, the Ford
Foundation, the Kavli Institute for Cosmological Physics, and University of
Chicago/Fermilab Strategic Collaborative Initiatives. The Holometer team
gratefully acknowledges the extensive support and contributions of Bradford
Boonstra, Benjamin Brubaker, Marcin Burdzy, Herman Cease, Tim Cunneen, Steve
Dixon, Bill Dymond, Valera Frolov, Jose Gallegos, Hank Glass, Emily Griffith,
Hartmut Grote, Gaston Gutierrez, Evan Hall, Sten Hansen, Young-Kee Kim, Mark
Kozlovsky, Dan Lambert, Scott McCormick, Erik Ramberg, Doug Rudd, Geoffrey
Schmit, Alex Sippel, Jason Steffen, Sali Sylejmani, David Tanner, Jim Volk,
William Wester, and James Williams towards the design and construction of the
apparatus.

\end{acknowledgments}

\bibliography{holobib} 
\end{document}